\documentclass[12pt]{article}
\usepackage{amssymb,amscd,array}
\catcode `\@=11
\@addtoreset{equation}{section}

\newtheorem{thm}{Theorem}[section]

\newtheorem{lemma}[thm]{Lemma}
\newtheorem{prop}[thm]{Proposition}
\newtheorem{cor}[thm]{Corollary}

\def\qed{\blacksquare}
\newcommand{\be}{\begin{equation}}
\newcommand{\ee}{\end{equation}}
\newcommand{\bea}{\begin{eqnarray}}
\newcommand{\eea}{\end{eqnarray}}
\newcommand{\R}{\mathbb{R}}

\newcommand{\C}{\mathbb{C}}

\begin{document}
\begin{titlepage}

\begin{center}
{\bf \Large{Factorization Formulas for Tree Amplitudes \\}}
\end{center}
\vskip 1.0truecm
\centerline{D. R. Grigore, 
\footnote{e-mail: grigore@theory.nipne.ro}}
\vskip5mm
\centerline{Department of Theoretical Physics,}
\centerline{Institute for Physics and Nuclear Engineering ``Horia Hulubei"}
\centerline{Bucharest-M\u agurele, P. O. Box MG 6, ROM\^ANIA}

\vskip 2cm
\bigskip \nopagebreak
\begin{abstract}
\noindent
We present a coordinate space version of the factorization formula for the connected tree part of the chronological products.
We consider a general framework and then we apply it for the QCD case.
\end{abstract}
\end{titlepage}

\section{Introduction}

The most natural way to arrive at the Bogoliubov axioms of perturbative quantum field theory (pQFT) is by
analogy with non-relativistic quantum mechanics \cite{Gl}, \cite{H}, \cite{D}. So we start from Bogoliubov axioms \cite{BS}, \cite{EG} 
as presented in \cite{algebra}; 
for every set of monomials 
$ 
A_{1}(x_{1}),\dots,A_{n}(x_{n}) 
$
in some jet variables (associated to some classical field theory) one associates the operator-valued distributions
$ 
T^{A_{1},\dots,A_{n}}(x_{1},\dots,x_{n})
$  
called chronological products; it will be convenient to use another notation: 
$ 
T(A_{1}(x_{1}),\dots,A_{n}(x_{n})). 
$ 

The Bogoliubov axioms, presented in Section \ref{Bogoliubov} express essentially some properties of the scattering matrix understood as a 
formal perturbation
series with the ``coefficients" the chronological products: (1) (sqew)symmetry properties in the entries 
$ 
A_{1}(x_{1}),\dots,A_{n}(x_{n}) 
$;
(2) Poincar\'e invariance; (3) causality; (4) unitarity; (5) the ``initial condition" which says that
$
T(A(x)) 
$
is a Wick monomial. So we need some basic notions on free fields and Wick monomials which will be presented in Section \ref{wick prod} 
also following \cite{algebra}.  One can supplement these axioms by requiring (6) power counting; (7) Wick expansion property. 
It is a highly non-trivial problem to find solutions for the Bogoliubov axioms, even in the simplest case of a real scalar field. 

There are, at least to our knowledge, tree rigorous ways to do that; for completeness we remind them following \cite{ano-free}:
(a) {\it Hepp axioms} \cite{H};
(b) {\it Polchinski flow equations} \cite{P}, \cite{S};
(c) {\it the causal approach} due to Epstein and Glaser \cite{EG}, \cite{Gl} which we prefer. It is a recursive procedure for the basic objects
$ 
T(A_{1}(x_{1}),\dots,A_{n}(x_{n}))
$
and reduces the induction procedure to a distribution splitting of some distributions with causal support.  
In an equivalent way, one can reduce the induction procedure to the process of extension of distributions \cite{PS}. 
An equivalent point of view uses retarded products \cite{St1} instead of chronological products. For gauge models one has to deal with non-physical 
fields (the so-called ghost fields) and impose a supplementary axiom (8) namely  gauge invariance, which guarantees that the 
physical states are left invariant by the chronological products.

In this paper we will prove some factorization properties of connected tree contributions of the chronological products. Such 
sort of factorization formulas have appeared a long time ago \cite{BG2}, \cite{Di}, \cite{BCFW} in momentum space. For a pedagogical 
exposition see for instance \cite{EKMZ} and \cite{HP}.
We give a purely combinatorial proof of some formulas of such a type in coordinate space. The framework is quite general: it works for any
tri-linear interaction Lagrangian. In the last Section \ref{ym} we will consider the particular case of QCD and try to obtain the 
formulas of \cite{BG2} stripping the Feynman amplitudes of the color factors. We present a quite general method and we use only well
defined mathematical objects: the chronological products. A similar analysis appears in \cite{E}.
\newpage

\section{Wick Products\label{wick prod}}

We follow the formalism from \cite{algebra}. We consider a classical field theory on the Minkowski space
$
{\cal M} \simeq \R^{4}
$
(with variables
$
x^{\mu}, \mu = 0,\dots,3
$
and the metric $\eta$ with 
$
diag(\eta) = (1,-1,-1,-1)
$)
described by the Grassmann manifold 
$
\Xi_{0}
$
with variables
$
\xi_{a}, a \in {\cal A}
$
(here ${\cal A}$ is some index set) and the associated jet extension
$
J^{r}({\cal M}, \Xi_{0}),~r \geq 1
$
with variables 
$
x^{\mu},~\xi_{a;\mu_{1},\dots,\mu_{n}},~n = 0,\dots,r;
$
we denote generically by
$
\xi_{p}, p \in P
$
the variables corresponding to classical fields and their formal derivatives and by
$
\Xi_{r}
$
the linear space generated by them. The variables from
$
\Xi_{r}
$
generate the algebra
$
{\rm Alg}(\Xi_{r})
$
of polynomials.

To illustrate this, let us consider a real scalar field in Minkowski space ${\cal M}$. The first jet-bundle extension is
$$
J^{1}({\cal M}, \R) \simeq {\cal M} \times \R \times \R^{4}
$$
with coordinates 
$
(x^{\mu}, \phi, \phi_{\mu}),~\mu = 0,\dots,3.
$

If 
$
\varphi: \cal M \rightarrow \R
$
is a smooth function we can associate a new smooth function
$
j^{1}\varphi: {\cal M} \rightarrow J^{1}(\cal M, \R) 
$
according to 
$
j^{1}\varphi(x) = (x^{\mu}, \varphi(x), \partial_{\mu}\varphi(x)).
$

For higher order jet-bundle extensions we have to add new real variables
$
\phi_{\{\mu_{1},\dots,\mu_{r}\}}
$
considered completely symmetric in the indexes. For more complicated fields, one needs to add supplementary indexes to
the field i.e.
$
\phi \rightarrow \phi_{a}
$
and similarly for the derivatives. The index $a$ carries some finite dimensional representation of
$
SL(2,\C)
$
(Poincar\'e invariance) and, maybe a representation of other symmetry groups. 
In classical field theory the jet-bundle extensions
$
j^{r}\varphi(x)
$
do verify Euler-Lagrange equations. To write them we need the formal derivatives defined by
\be
d_{\nu}\phi_{\{\mu_{1},\dots,\mu_{r}\}} \equiv \phi_{\{\nu,\mu_{1},\dots,\mu_{r}\}}.
\ee

We suppose that in the algebra 
$
{\rm Alg}(\Xi_{r})
$
generated by the variables 
$
\xi_{p}
$
there is a natural conjugation
$
A \rightarrow A^{\dagger}.
$
If $A$ is some monomial in these variables, there is a canonical way to associate to $A$ a Wick 
monomial: we associate to every classical field
$
\xi_{a}, a \in {\cal A}
$
a quantum free field denoted by
$
\xi^{\rm quant}_{a}(x), a \in {\cal A}
$
and determined by the $2$-point function
\be
<\Omega, \xi^{\rm quant}_{a}(x), \xi^{\rm quant}_{b}(y) \Omega> = - i~D^{(+)}(\xi_{a}(x),\xi_{b}(y))\times {\bf 1}.
\label{2-point}
\ee
Here 
\be
D_{ab}(x - y) \equiv D(\xi_{a}(x),\xi_{b}(y))
\ee
is the causal Pauli-Jordan distribution associated to the two fields; it is (up to some numerical factors) a polynomial
in the derivatives applied to the Pauli-Jordan distribution. We understand by 
$
D^{(\pm)}_{ab}(x)
$
the positive and negative parts of
$
D_{ab}(x)
$.
The $n$-point functions for
$
n \geq 3
$
are obtained assuming that the truncated Wightman functions are null: see \cite{BLOT}, relations (8.74) and (8.75) and proposition 8.8
from there. The definition of these truncated Wightman functions involves the Fermi parities
$
|\xi_{p}|
$
of the fields
$
\xi_{p}, p \in P.
$

Afterwards we define
$$
\xi^{\rm quant}_{a;\mu_{1},\dots,\mu_{n}}(x) \equiv \partial_{\mu_{1}}\dots \partial_{\mu_{n}}\xi^{\rm quant}_{a}(x), a \in {\cal A}
$$
which amounts to
\be
D( \xi_{a;\mu_{1}\dots\mu_{m}}(x), \xi_{b;\nu_{1}\dots\nu_{n}}(y) ) =
(-1)^{n}~i~\partial_{\mu_{1}}\dots \partial_{\mu_{m}}\partial_{\nu_{1}}\dots \partial_{\nu_{n}}D_{ab}(x - y )\times {\bf 1}.
\label{2-point-der}
\ee
More sophisticated ways to define the free fields involve the GNS construction. 

The free quantum fields are generating a Fock space 
$
{\cal F}
$
in the sense of the Borchers algebra: formally it is generated by states of the form
$
\xi^{\rm quant}_{a_{1}}(x_{1})\dots \xi^{\rm quant}_{a_{n}}(x_{n})\Omega
$
where 
$
\Omega
$
the vacuum state.
The scalar product in this Fock space is constructed using the $n$-point distributions and we denote by
$
{\cal F}_{0} \subset {\cal F}
$
the algebraic Fock space.

One can prove that the quantum fields are free, i.e.
they verify some free field equation; in particular every field must verify Klein Gordon equation for some mass $m$
\be
(\square + m^{2})~\xi^{\rm quant}_{a}(x) = 0
\ee
and it follows that in momentum space they must have the support on the hyperboloid of mass $m$. This means that 
they can be split in two parts
$
\xi^{\rm quant (\pm)}_{a}
$
with support on the upper (resp. lower) hyperboloid of mass $m$. We convene that 
$
\xi^{\rm quant (+)}_{a} 
$
resp.
$
\xi^{\rm quant (-)}_{a} 
$
correspond to the creation (resp. annihilation) part of the quantum field. The expressions
$
\xi^{\rm quant (+)}_{p} 
$
resp.
$
\xi^{\rm quant (-)}_{p} 
$
for a generic
$
\xi_{p},~ p \in P
$
are obtained in a natural way, applying partial derivatives. For a general discussion of this method of constructing free fields, see 
ref. \cite{BLOT} - especially prop. 8.8.
The Wick monomials are leaving invariant the algebraic Fock space.
The definition for the Wick monomials is contained in the following Proposition.

\begin{prop}
The operator-valued distributions
$
N(\xi_{q_{1}}(x_{1}),\dots,\xi_{q_{n}}(x_{n}))
$
are uniquely defined by:

\be
N(\xi_{q_{1}}(x_{1}),\dots,\xi_{q_{n}}(x_{n}))\Omega = \xi_{q_{1}}^{(+)}(x_{1})\dots  \xi_{q_{n}}^{(+)}(x_{n})\Omega
\ee

\bea
[ \xi_{p}(y), N(\xi_{q_{1}}(x_{1}),\dots,\xi_{q_{n}}(x_{n})) ] = 
\nonumber\\
- i~\sum_{m=1}^{n} \prod_{l <m} (-1)^{|\xi_{p}||\xi_{q_{l}}|}~D_{pq_{m}}(y - x_{m})~N(\xi_{q_{1}}(x_{1}),\dots,\hat{m},\dots,\xi_{q_{n}}(x_{n}))
\eea

\be
N(\emptyset) = I.
\ee

The expression
$
N(\xi_{q_{1}}(x_{1}),\dots,\xi_{q_{n}}(x_{n}))
$
is (graded) symmetrical in the arguments.
\end{prop}

The expression
$
N(\xi_{q_{1}}(x_{1}),\dots,\xi_{q_{n}}(x_{n}))
$
are called {\it Wick monomials}. There is an alternative definition based on the splitting of the fields into the creation and annihilation 
part for which we refer to \cite{algebra}.

It is a non-trivial result of Wightman and G\aa rding \cite{WG} that in
$
N(\xi_{q_{1}}(x_{1}),\dots,\xi_{q_{n}}(x_{n}))
$
one can collapse all variables into a single one and still gets an well-defined expression: 
\begin{prop}
The expressions
\be
W_{q_{1},\dots,q_{n}}(x) \equiv N(\xi_{q_{1}}(x),\dots,\xi_{q_{n}}(x))
\ee
are well-defined. They verify:

\be
W_{q_{1},\dots,q_{n}}(x) \Omega = \xi_{q_{1}}^{(+)}(x)\dots  \xi_{q_{n}}^{(+)}(x)\Omega
\ee

\bea
[ \xi^{(\epsilon)}_{p}(y), W_{q_{1},\dots,q_{n}}(x)  ] = 
- i~\sum_{m=1}^{n} \prod_{l <m} (-1)^{|\xi_{p}||\xi_{q_{l}}|}~D_{pq_{m}}^{(-\epsilon)}(y - x_{m})~W_{q_{1},\dots,\hat{m},\dots,q_{n}}(x)
\eea

\be
W(\emptyset) = I.
\ee
\end{prop}
We call expressions of the type
$
W_{q_{1},\dots,q_{n}}(x) 
$
{\it Wick monomials}.  By
\be
|W| \equiv \sum_{l=1}^{n} |\xi_{q_{l}}|
\ee
we mean the Fermi number of $W$. We define the derivative
\be
{\partial \over \partial \xi_{p}}W_{q_{1},\dots,q_{n}}(x) \equiv 
\sum_{s=1}^{n}~\prod_{l < s}~(-1)^{|\xi_{p}||\xi_{q_{l}}|}~\delta_{pq_{s}}~W_{q_{1},\dots,\hat{q_{s}},\dots,q_{n}}(x)
\ee
and we have a generalization of the preceding Proposition.
\begin{prop}
Let
$
W_{j} = W_{q^{(j)}_{1},\dots,q^{(j)}_{r_{j}}},~j = 1,\dots,n
$
be Wick monomials. Then the expression
$
N(W_{1}(x_{1}),\dots,W_{n}(x_{n}))
$
is well-defined through

\be
N(W_{1}(x_{1}),\dots,W_{n}(x_{n}))\Omega = \prod_{j=1}^{n} \prod_{l=1}^{r_{j}} \xi_{q^{(j)}_{l}}^{(+)}(x_{j})\Omega
\ee

\bea
[ \xi_{p}(y), N(W_{1}(x_{1}),\dots,W_{n}(x_{n}))  ] = 
\nonumber\\
- i~\sum_{m=1}^{n} \prod_{l <m} (-1)^{|\xi_{p}||W_{l}|}~\sum_{q}~D_{pq}(y - x_{m})~
N(W_{1}(x_{1}),\dots,{\partial \over \partial \xi_{q}}W_{m}(x_{m}),\dots,W_{n}(x_{n}))
\label{normal}
\eea

\be
N(W_{1}(x_{1}),\dots,W_{n}(x_{n}),{\bf 1}) = N(W_{1}(x_{1}),\dots,W_{n}(x_{n})) 
\ee

\be
N(W(x)) = W(x).
\ee

The expression
$
N(W_{1}(x_{1}),\dots,W_{n}(x_{n}))
$
is symmetric (in the Grassmann sense) in the entries
$
W_{1}(x_{1}),\dots,W_{n}(x_{n}).
$
\label{N}
\end{prop}

One can prove that 
\be
[ N(A(x)), N(B(y)) ] = 0,\quad (x - y)^{2} < 0
\ee
where by
$
[ \cdot,\cdot]
$
we mean the graded commutator. This is the most simple case of causal support property.
Now we are ready for the most general setting. We define for any monomial
$
A \in {\rm Alg}(\Xi_{r})
$
the derivation
\be
\xi \cdot A \equiv (-1)^{|\xi| |A|}~{\partial \over \partial \xi}A
\label{derivative}
\ee
for all
$
\xi \in \Xi_{r}.
$
Here 
$|A|$ 
is the Fermi parity of $A$ and we consider the left derivative in the Grassmann sense. So, for 
the moment, the product $\cdot$ is defined as an map
$
\Xi_{r} \times {\rm Alg}(\Xi_{r}) \rightarrow {\rm Alg}(\Xi_{r}).
$
An expression
$
E(A_{1}(x_{1}),\dots,A_{n}(x_{n}))
$
is called {\it of Wick type} iff verifies:

\bea
[ \xi_{p}(y), E(A_{1}(x_{1}),\dots,A_{n}(x_{n}))  ] = 
\nonumber\\
- i~\sum_{m=1}^{n} \prod_{l \leq m} (-1)^{|\xi_{p}||A_{l}|}~\sum_{q}~D_{pq}(y - x_{m})~
E(A_{1}(x_{1}),\dots,\xi_{q}\cdot A_{m}(x_{m}),\dots,A_{n}(x_{n}))
\label{comm-wick}
\eea

\be
E(A_{1}(x_{1}),\dots,A_{n}(x_{n}),{\bf 1}) = E(A_{1}(x_{1}),\dots,A_{n}(x_{n})) 
\ee

\be
E(1) = {\bf 1}.
\ee

The expression
$
N(W_{1}(x_{1}),\dots,W_{n}(x_{n}))
$
from  Proposition \ref{N} is of Wick type. Then we easily have:
\begin{prop}
If
$
E(A_{1}(x_{1}),\dots,A_{k}(x_{k}))
$
and
$
F(A_{k+1}(x_{1}),\dots,A_{n}(x_{n}))
$
are expressions of Wick type, then
$
E(A_{1}(x_{1}),\dots,A_{k}(x_{k}))~F(A_{k+1}(x_{1}),\dots,A_{n}(x_{n}))
$
is also an expression of Wick type.
\end{prop}

Now we formulate Wick theorem. First we extend the product (\ref{derivative}) to more factors through iteration:
\be
(\xi\eta)\cdot A \equiv \xi\cdot (\eta\cdot A),\quad \xi,\eta \in \Xi_{0}
\ee
and $A$ an arbitrary monomial. So now $\cdot$ is a map
$
{\rm Alg}(\Xi_{r}) \times {\rm Alg}(\Xi_{r}) \rightarrow {\rm Alg}(\Xi_{r})
$
In particular it makes sense to consider expressions of the type
$
B\cdot A
$
where $A$ and $B$ are both monomials. One gets something non-null if $B$ is a {\it submonomial} of $A$. One easily derives that 
\be
A\cdot A = C(A) {\bf 1}
\ee
where 
$
C(A)
$
is a numerical factor. Then we have:
\begin{thm} ({\bf Wick})
Let
$
E(A_{1}(x_{1}),\dots,A_{n}(x_{n}))
$
be an expression of Wick type. The following formula is true:
\bea
E(A_{1}(x_{1}),\dots,A_{n}(x_{n})) = 
\sum_{B \in \Xi}~\epsilon(B_{1},\dots,B_{n};A_{1},\dots,A_{n})
\nonumber\\
<\Omega, E(B_{1}(x_{1}),\dots,B_{n}(x_{n}))\Omega>~
N(B_{1}\cdot A_{1}(x_{1}),\dots,B_{n}\cdot A_{n}(x_{n}))
\label{wick-thm}
\eea
where 
$
B_{j}
$
are distinct Wick submonomials of
$
A_{j}
$
and
\be
\epsilon(B_{1},\dots,B_{n};A_{1},\dots,A_{n}) \equiv (-1)^{s}~\prod_{l=1}^{n}~C(B_{l})^{-1}
\ee
with 
\be
s \equiv \sum_{l=1}^{n} |B_{l}|~(\sum_{p=l+1}^{n}~(|A_{p}| + |B_{p}|) = \sum_{p=2}^{n}~(|A_{p}| + |B_{p}|)~(\sum_{l=1}^{p-1}~|B_{l}|). 
\ee
\end{thm}

In the same way we prove:
\begin{thm}
The following formula is true:
\bea
N(\xi_{p}(y),A_{1}(x_{1}),\dots,A_{n}(x_{n})) = \xi_{p}(y)~N(A_{1}(x_{1}),\dots,A_{n}(x_{n}))
\nonumber\\
+ i~\sum_{m=1}^{n} \prod_{l <m} (-1)^{|\xi_{p}||A_{l}|}~\sum_{q}~D^{(+)}_{pq}(y - x_{m})~
N(A_{1}(x_{1}),\dots,\xi_{q}\cdot A_{m}(x_{m}),\dots,A_{n}(x_{n}))
\eea
\end{thm}

\newpage
\section{Bogoliubov Axioms \label{Bogoliubov}}
Suppose the monomials
$
A_{1},\dots,A_{n} \in {\rm Alg}(\Xi_{r})
$
are self-adjoint:
$
A_{j}^{\dagger} = A_{j},~\forall j = 1,\dots,n
$
and of Fermi number
$
f_{i}.
$

The chronological products
$$ 
T(A_{1}(x_{1}),\dots,A_{n}(x_{n})) \equiv T^{A_{1},\dots,A_{n}}(x_{1},\dots,x_{n}) \quad n = 1,2,\dots
$$
are some distribution-valued operators leaving invariant the algebraic Fock space and verifying the following set of axioms:
\begin{itemize}
\item
{\bf Skew-symmetry} in all arguments:
\be
T(\dots,A_{i}(x_{i}),A_{i+1}(x_{i+1}),\dots,) =
(-1)^{f_{i} f_{i+1}} T(\dots,A_{i+1}(x_{i+1}),A_{i}(x_{i}),\dots)
\ee

\item
{\bf Poincar\'e invariance}: we have a natural action of the Poincar\'e group in the
space of Wick monomials and we impose that for all 
$g \in inSL(2,\C)$
we have:
\be
U_{g} T(A_{1}(x_{1}),\dots,A_{n}(x_{n})) U^{-1}_{g} =
T(g\cdot A_{1}(x_{1}),\dots,g\cdot A_{n}(x_{n}))
\label{invariance}
\ee
where in the right hand side we have the natural action of the Poincar\'e group on
$
\Xi
$.

Sometimes it is possible to supplement this axiom by other invariance
properties: space and/or time inversion, charge conjugation invariance, global
symmetry invariance with respect to some internal symmetry group, supersymmetry,
etc.
\item
{\bf Causality}: if 
$
y \cap (x + \bar{V}^{+}) = \emptyset
$
then we denote this relation by
$
x \succeq y
$.
Suppose that we have 
$x_{i} \succeq x_{j}, \quad \forall i \leq k, \quad j \geq k+1$.
then we have the factorization property:
\be
T(A_{1}(x_{1}),\dots,A_{n}(x_{n})) =
T(A_{1}(x_{1}),\dots,A_{k}(x_{k}))~~T(A_{k+1}(x_{k+1}),\dots,A_{n}(x_{n}));
\label{causality}
\ee

\item
{\bf Unitarity}: We define the {\it anti-chronological products} using a convenient notation introduced
by Epstein-Glaser, adapted to the Grassmann context. If 
$
X = \{j_{1},\dots,j_{s}\} \subset N \equiv \{1,\dots,n\}
$
is an ordered subset, we define
\be
T(X) \equiv T(A_{j_{1}}(x_{j_{1}}),\dots,A_{j_{s}}(x_{j_{s}})).
\ee
Let us consider some Grassmann variables
$
\theta_{j},
$
of parity
$
f_{j},  j = 1,\dots, n
$
and let us define
\be
\theta_{X} \equiv \theta_{j_{1}} \cdots \theta_{j_{s}}.
\ee
Now let
$
(X_{1},\dots,X_{r})
$
be a partition of
$
N = \{1,\dots,n\}
$
where
$
X_{1},\dots,X_{r}
$
are ordered sets. Then we define the (Koszul) sign
$
\epsilon(X_{1},\dots,X_{r})
$
through the relation
\be
\theta_{1} \cdots \theta_{n} = \epsilon(X_{1}, \dots,X_{r})~\theta_{X_{1}} \dots \theta_{X_{r}}
\ee
and the antichronological products are defined according to
\be
(-1)^{n} \bar{T}(N) \equiv \sum_{r=1}^{n} 
(-1)^{r} \sum_{I_{1},\dots,I_{r} \in Part(N)}
\epsilon(X_{1},\dots,X_{r})~T(X_{1})\dots T(X_{r})
\label{antichrono}
\ee
Then the unitarity axiom is:
\be
\bar{T}(N) = T(N)^{\dagger}.
\label{unitarity}
\ee
\item
{\bf The ``initial condition"}:
\be
T(A(x)) =  N(A(x)).
\ee

\item
{\bf Power counting}: We can also include in the induction hypothesis a limitation on the order of
singularity of the vacuum averages of the chronological products associated to
arbitrary Wick monomials
$A_{1},\dots,A_{n}$;
explicitly:
\be
\omega(<\Omega, T^{A_{1},\dots,A_{n}}(X)\Omega>) \leq
\sum_{l=1}^{n} \omega(A_{l}) - 4(n-1)
\label{power}
\ee
where by
$\omega(d)$
we mean the order of singularity of the (numerical) distribution $d$ and by
$\omega(A)$
we mean the canonical dimension of the Wick monomial $W$.

\item
{\bf Wick expansion property}: In analogy to (\ref{comm-wick}) we require
\bea
[ \xi_{p}(y), T(A_{1}(x_{1}),\dots,A_{n}(x_{n})) ]
\nonumber\\
= - i~\sum_{m=1}^{n}~\prod_{l \leq m} (-1)^{|\xi_{p}||A_{l}|}~\sum_{q}~D_{pq}(y - x_{m} )~
T(A_{1}(x_{1}),\dots,\xi_{q}\cdot A_{m}(x_{m}),\dots, A_{n}(x_{n}))
\nonumber\\
\label{wick}
\eea
\end{itemize}

Up to now, we have defined the chronological products only for self-adjoint Wick monomials 
$
W_{1},\dots,W_{n}
$
but we can extend the definition for Wick polynomials by linearity.

The construction of Epstein-Glaser is based on a recursive procedure \cite{EG}. 
We can derive from these axioms the following result \cite{Sto1}.
\begin{thm}
One can fix the causal products such that the following formula is true
\bea
T(\xi_{p}(y), A_{1}(x_{1}),\dots,A_{n}(x_{n}))
\nonumber\\
= - i~\sum_{m=1}^{n}~\prod_{l \leq m} (-1)^{|\xi_{p}||A_{l}|}~\sum_{q}~D^{F}_{pq}(y - x_{m} )~
T(A_{1}(x_{1}),\dots,\xi_{q}\cdot A_{m}(x_{m}),\dots, A_{n}(x_{n}))
\nonumber\\
+ \xi^{(+)}_{p}(y)~T(A_{1}(x_{1}),\dots, A_{n}(x_{n}))
+ \prod_{l \leq n}~(-1)^{|\xi_{p}| f_{l}}~T(A_{1}(x_{1}),\dots, A_{n}(x_{n}))~\xi^{(-)}_{p}(y)
\label{linear}
\eea
where 
$
D^{F}_{pq}
$
is a Feynman propagator associated to the causal distribution
$
D_{pq}
$.
\end{thm}
Some times (\ref{linear}) - or  variants of it - is called the {\bf equation of motion} axiom \cite{DF}.

\newpage
\section{Tree Contributions\label{tree}}

For simplicity we assume that the jet space is generated by scalars variables
$
\xi_{a}, a \in {\cal A}. 
$

If $A$ is a  monomial in the jet variables, then we denote by
$
\rho(A)
$
the number of factors. Then we have:
\begin{prop}
Let 
$
A_{1},\dots,A_{n}
$
such that
\be
\sum_{j=1}^{n} \rho(A_{j}) = 2 n - 2.
\ee
Then the chronological product
$
T^{\rm conn}(A_{1}(x_{1}),\dots,A_{n}(x_{n}))
$
is a $c$-number.
\label{t-conn}
\end{prop}
{\bf Proof:}
We use the topological relations valid for the connected part of the chronological products:
\bea
L = I - n + 1
\nonumber\\
\sum_{j=1}^{n} \rho(A_{j}) = 2 I + B
\eea
where $L$ is the number of loops, $I$ is the number of internal lines and $B$ the number of external lines. If we eliminate $I$ 
and use the relation from the statement, we end up with
\bea
2 L + B = 0
\nonumber
\eea
so we have
$
L = 0, B = 0.
$
In particular, from 
$
B = 0
$
it follows that we do not have external lines so 
$
T^{\rm conn}(A_{1}(x_{1}),\dots,A_{n}(x_{n}))
$
is a $c$-number.
$\qed$
\begin{cor}
In the preceding conditions, suppose that we have 
$
p_{1}
$
entries 
$A$
with
$
\rho(A) = 1, 
$
$
p_{2}
$
entries with
$
\rho(A) = 2 
$
and
$
p_{3}
$
entries with
$
\rho(A) = 3.
$
Then we have
\bea
p_{1} = p_{3} + 2
\nonumber\\
p_{2} = n - 2 - 2 p_{3}
\eea
so in particular we must have
$
n \geq 2 (p_{3} + 1).
$
\end{cor}
\newpage
\begin{prop}
Let 
$
A_{2},\dots,A_{n-1}
$
with
$
\rho(A_{j}) = 2, \forall j.
$
Then the following formula is valid
\bea
T^{\rm conn}(\xi_{a_{1}}(x_{1}), A_{2}(x_{2}),\dots,A_{n-1}(x_{n-1}),\xi_{a_{n}}(x_{n})) =
\nonumber\\
( - i)^{n - 1}~(n - 2)!~{\cal S} [ D^{F}(\xi_{a_{1}}(x_{1}), \xi_{a_{2}}(x_{2}))~
\nonumber\\
\prod_{j= 2}^{n - 2} D^{F}(\xi_{a_{j}}\cdot A_{j}(x_{j}), \xi_{a_{j + 1}}(x_{j + 1}))~
D^{F}(\xi_{a_{n - 1}}\cdot A_{n - 1}(x_{n - 1}), \xi_{a_{n}}(x_{n})) ] \quad ( n \geq 3)
\eea
where 
$
{\cal S}
$
symmetrizes in 
$
A_{2}(x_{2}),\dots,A_{n-1}(x_{n-1}).
$
\end{prop}
{\bf Proof:} By induction on $n$. For 
$
n = 3
$
this follows immediately from the field equation property; then the induction goes rather easily using again the field equation property. 
We remark that, because the monomials 
$
A
$
are bilinear, then the formal derivative
$
\xi\cdot A
$
is a linear expression in the fields, so the expression
$
D^{F}(\xi\cdot A(x), \eta(y))
$
makes sense.
$\qed$

From the preceding Proposition we have
\begin{cor}
The preceding formula can be written in the following form 
\bea
T^{\rm conn}(\xi_{a_{1}}(x_{1}), A_{2}(x_{2}),\dots,A_{n-1}(x_{n-1}),\xi_{a_{n}}(x_{n})) =
\nonumber\\
\sum_{p= 2}^{n - 1}~ C_{n - 3}^{p - 2}~{\cal S} [ T^{\rm conn}(\xi_{a_{1}}(x_{1}), A_{2}(x_{2}),\dots,A_{p-1}(x_{p-1}),\xi_{a_{p}}(x_{p}))
\nonumber\\
T^{\rm conn}(\xi_{a_{p}}\cdot A_{p}(x_{p}),A_{p + 1}(x_{p + 1}),\dots, A_{n - 1}(x_{n - 1}),\xi_{a_{n}}(x_{n})) ]
\quad ( n \geq 3)
\eea
\end{cor}
We have our first form of a factorization formula. However, we would want to get rid of the combinatorial factors.

The first step is:
\begin{cor}
Suppose that
$$
A_{p} =  {1\over 2}~g_{bc}^{(p)}~\xi_{b}\xi_{c},\quad g_{bc}^{(p)} = b \leftrightarrow c.
$$
Then, the preceding formula can be written in the following form 
\bea
T^{\rm conn}(\xi_{a_{1}}(x_{1}), A_{2}(x_{2}),\dots,A_{n-1}(x_{n-1}),\xi_{a_{n}}(x_{n})) =
\nonumber\\
{1 \over n - 2}~\sum_{p= 2}^{n - 1}~g_{bc}^{(p)}~C_{n - 3}^{p - 2}~{\cal S}_{(p)}
[ T^{\rm conn}(\xi_{a_{1}}(x_{1}), A_{2}(x_{2}),\dots,A_{p-1}(x_{p-1}),\xi_{b}(x_{p}))
\nonumber\\
T^{\rm conn}(\xi_{c}(x_{p}),A_{p + 1}(x_{p + 1}),\dots, A_{n - 1}(x_{n - 1}),\xi_{a_{n}}(x_{n})) ]
\quad ( n \geq 3)
\eea
where
$
{\cal S}_{(p)}
$
symmetrizes in
$
A_{2}(x_{2}),\dots,A_{p-1}(x_{p-1}), A_{p+1}(x_{p +1}),\dots,A_{n-1}(x_{n-1}).
$
\end{cor}
{\bf Proof:} We use
$
{\cal S} = {1 \over n - 2}~\sum_{p= 2}^{n - 1}~{\cal S}_{(p)}
$
and
$
\xi_{b}\cdot A_{p} = g_{bc}^{(p)}~\xi_{c}.
$
$\qed$

Next we have:
\begin{prop}
Let
$
A_{1},\dots,A_{n}
$
with
$
\rho(A_{j}) = 2, \forall j.
$
Then
\bea
T^{\rm conn}(\xi_{a_{1}}(z_{1}), \xi_{a_{2}}(z_{2}), A_{1}(x_{1}),\dots,A_{n}(x_{n})) =
\nonumber\\
{1 \over n}~\sum_{p = 1}^{n}~g_{bc}^{(p)}~\sum_{I_{1},I_{2} \in Part(N_{p})}
T^{\rm conn}(\xi_{a_{1}}(z_{1}), \xi_{b}(x_{p}), I_{1})~ 
T^{\rm conn}(\xi_{a_{2}}(z_{2}), \xi_{c}(x_{p}), I_{2})
\eea
Here 
$
N_{p} \equiv \{1,\dots,n\} - \{p\}
$
and for 
$
I = \{i_{1},\dots,i_{l}\},~ p \not\in I
$
we denote
\be
T^{\rm conn}(\xi_{a}(z), \xi_{b}(x_{p}), I) \equiv
T^{\rm conn}(\xi_{a}(z), \xi_{b}(x_{p}), A_{i_{1}}(x_{i_{1}}),\dots, A_{i_{l}}(x_{i_{l}})).
\ee
Equivalently
\bea
T^{\rm conn}(\xi_{a_{1}}(z_{1}), \xi_{a_{2}}(z_{2}), A_{1}(x_{1}),\dots,A_{n}(x_{n})) =
\nonumber\\
{1 \over n}~\sum_{I_{0},I_{1},I_{2} \in Part(N), |I_{0}| = 1}~g_{bc}^{I_{0}}~
T^{\rm conn}(\xi_{a_{1}}(z_{1}), \xi_{b}(x_{I_{0}}), I_{1})~ 
T^{\rm conn}(\xi_{a_{2}}(z_{2}), \xi_{c}(x_{I_{0}}), I_{2})
\eea
where if
$
I_{0} = \{p\}
$
we denote
$
g_{bc}^{I_{0}} \equiv g_{bc}^{(p)}
$
and 
$
x_{I_{0}} \equiv x_{p}.
$
From the last formula the symmetry property at
$
\xi_{a_{1}}(z_{1}) \leftrightarrow \xi_{a_{2}}(z_{2})
$
is obvious, as it is the symmetry in
$
A_{1}(x_{1}),\dots,A_{n}(x_{n}).
$
\end{prop}
{\bf Proof:} We consider the last formula from the statement. The symmetry at
$
\xi_{a_{1}}(z_{1}) \leftrightarrow \xi_{a_{2}}(z_{2})
$
follows if we use the redefinitions
$
I_{1} \leftrightarrow I_{2},~b \leftrightarrow c.
$
For the symmetry at
$
A_{j}(x_{j}) \leftrightarrow A_{k}(x_{k})
$
(with 
$
j,k
$
fixed), we split the sum in three contributions: 
a) $j, k \in I_{1}$ (or  $j, k \in I_{2}$);
b) $j \in I_{1},k \in I_{2}$;
c) $p = j$ of $p = k$.
The first two contributions are obviously invariant to 
$
A_{j}(x_{j}) \leftrightarrow A_{k}(x_{k})
$
and the two contributions from case c) are mapped one into each other.

Now we prove that the last formula from the statement coincides with the formula from the preceding proposition. Because of the 
symmetry property in
$
A_{1}(x_{1}),\dots,A_{n}(x_{n})
$
just proved we can introduce the symmetrization operator
${\cal S}$
without changing anything:
\bea
T^{\rm conn}(\xi_{a_{1}}(z_{1}), \xi_{a_{2}}(z_{2}), A_{1}(x_{1}),\dots,A_{n}(x_{n})) =
\nonumber\\
{1 \over n}~\sum_{I_{0},I_{1},I_{2} \in Part(N), |I_{0}| = 1}~g_{bc}^{I_{0}}~{\cal S}
[ T^{\rm conn}(\xi_{a_{1}}(z_{1}), \xi_{b}(x_{I_{0}}), I_{1})~ 
T^{\rm conn}(\xi_{a_{2}}(z_{2}), \xi_{c}(x_{I_{0}}), I_{2}) ]
\nonumber
\eea
Now we consider all contributions with
$
|I_{1}| = p - 1.
$
There are
$
C_{n - 1}^{p - 1}
$
such contributions and all are equal because of the presence of the symmetrization operator
${\cal S}$.
So, we can take
$
I_{1} = \{1,\dots,p - 1\},~I_{2} = \{p + 1,\dots,n\}
$
and
$
I_{0} = \{p\}.
$
The formula from the preceding proposition follows.
$\qed$

Next we have
\begin{prop}
The $n$ contributions from the preceding proposition are equal. It follows that we have
\bea
T^{\rm conn}(\xi_{a_{1}}(z_{1}), \xi_{a_{2}}(z_{2}), A_{1}(x_{1}),\dots,A_{n}(x_{n})) =
\nonumber\\
g_{bc}^{(1)}~\sum_{I_{1},I_{2} \in Part(\{2,\dots,n\})}
T^{\rm conn}(\xi_{a_{1}}(z_{1}), \xi_{b}(x_{1}), I_{1})~ 
T^{\rm conn}(\xi_{a_{2}}(z_{2}), \xi_{c}(x_{1}), I_{2})
\eea
\label{A(0,n)}
\end{prop}
{\bf Proof:} By induction. For 
$
n = 2
$
the equality follows easily. We suppose that we have the equality for 
$
2,\dots,n - 1~(n \geq 3)
$
and consider the case $n$. It is sufficient to prove the equality of the first two contributions. In the expression
\bea
E_{1} \equiv g_{bc}^{(1)}~\sum_{I_{1},I_{2} \in Part(\{2,\dots,n\})}
T^{\rm conn}(\xi_{a_{1}}(z_{1}), \xi_{b}(x_{1}), I_{1})~ 
T^{\rm conn}(\xi_{a_{2}}(z_{2}), \xi_{c}(x_{1}), I_{2})
\nonumber
\eea
we have two types of terms: 
a) $2 \in I_{1}$,
b) $2 \in I_{2}$;
we apply in both contributions the induction hypothesis. We use the same idea in the contribution
\bea
E_{2} \equiv g_{bc}^{(2)}~\sum_{I_{1},I_{2} \in Part(\{1,3,\dots,n\})}
T^{\rm conn}(\xi_{a_{1}}(z_{1}), \xi_{b}(x_{2}), I_{1})~ 
T^{\rm conn}(\xi_{a_{2}}(z_{2}), \xi_{c}(x_{2}), I_{2})
\nonumber
\eea
and we obtain
$
E_{1} = E_{2}.
$
$\qed$

Next we introduce in the game a Wick monomial with tree factors.
\begin{prop}
Let
$
A_{1},\dots,A_{n}
$
with
$
\rho(A_{j}) = 2, \forall j
$
and
\be
B = {1\over 3!}~g_{b_{1}b_{2}b_{3}}~\xi_{b_{1}}~\xi_{b_{2}}~\xi_{b_{3}}
\label{tri}
\ee
with
$
g_{b_{1}b_{2}b_{3}}
$
completely symmetric. Then
\bea
T^{\rm conn}(\xi_{a_{1}}(z_{1}), \xi_{a_{2}}(z_{2}), \xi_{a_{3}}(z_{3}), A_{1}(x_{1}),\dots,A_{n}(x_{n}), B(y)) =
\nonumber\\
g_{b_{1}b_{2}b_{3}}~\sum_{I_{1},I_{2},I_{3} \in Part(\{1,\dots,n\})}
T^{\rm conn}(\xi_{a_{1}}(z_{1}), \xi_{b_{1}}(y), I_{1})~ 
\nonumber\\
T^{\rm conn}(\xi_{a_{2}}(z_{2}), \xi_{b_{2}}(y), I_{2})~
T^{\rm conn}(\xi_{a_{3}}(z_{3}), \xi_{b_{3}}(y), I_{3}).
\eea
The symmetry in
$
\xi_{a_{j}}(z_{j})
$
and in
$
A_{1}(x_{1}),\dots,A_{n}(x_{n})
$
is manifest.
\label{B(1,n)}
\end{prop}
{\bf Proof:} By induction on $n$. For 
$
n = 0
$
we use the equation of motion axiom and obtain the formula. We consider the formula valid for 
$
1,\dots,n - 1
$
and we have for $n$ with the equation of motion axiom:
\bea
T^{\rm conn}(\xi_{a_{1}}(z_{1}), \xi_{a_{2}}(z_{2}), \xi_{a_{3}}(z_{3}), A_{1}(x_{1}),\dots,A_{n}(x_{n}), B(y)) =
- i~\sum_{j=1}^{n}~D^{F}(\xi_{a_{1}}(z_{1}),\xi_{d}(x_{j}))~
\nonumber\\
T^{\rm conn}(\xi_{a_{2}}(z_{2}), \xi_{a_{3}}(z_{3}), A_{1}(x_{1}),\dots,\xi_{d}\cdot A_{j}(x_{j}), \dots,A_{n}(x_{n}), B(y))
\nonumber\\
- i~\sum_{j=1}^{n}~D^{F}(\xi_{a_{1}}(z_{1}),\xi_{d}(y))~
\nonumber\\
T^{\rm conn}(\xi_{a_{2}}(z_{2}), \xi_{a_{3}}(z_{3}), A_{1}(x_{1}),\dots, A_{n}(x_{n}), \xi_{d}\cdot B(y))
\eea
In the first contribution we apply the induction hypothesis and the second contribution we use the preceding proposition.
$\qed$

Next we have:
\begin{prop}
In the preceding conditions we have
\bea
T^{\rm conn}(\xi_{a_{1}}(z_{1}), \xi_{a_{2}}(z_{2}), \xi_{a_{3}}(z_{3}), A_{1}(x_{1}),\dots,A_{n}(x_{n}), B(y)) =
3~g_{bc}^{(1)}
\nonumber\\
\sum_{I_{1},I_{2} \in Part(\{2,\dots,n\})}~{\cal S}
[ T^{\rm conn}(\xi_{a_{1}}(z_{1}), \xi_{b}(x_{1}), I_{1})~ 
T^{\rm conn}(\xi_{a_{2}}(z_{2}), \xi_{a_{3}}(z_{3}), \xi_{c}(x_{1}), I_{2}, B(y)) ]
\eea
where 
$
{\cal S}
$
symmetrizes in
$
\xi_{a_{j}}(z_{j}),~j = 1,2,3.
$
We have similar formulas with
$
g_{bc}^{(1)} \rightarrow g_{bc}^{(j)}~(\forall j).
$
\label{A(1,n)}
\end{prop}
{\bf Proof:} In the right hand side of the formula from the statement we consider three contributions corresponding to 
$
1 \in I_{1}, 1 \in I_{2}, 1 \in I_{3}
$
and use the preceding proposition. Summing the three contributions we get the left hand side of the formula from the statement.
$\qed$

Finally we have the main result.
\begin{thm}
In the preceding conditions the following formula is true for 
$
m \geq 1
$
\bea
T^{\rm conn}(\xi_{a_{1}}(z_{1}), \dots, \xi_{a_{m+2}}(z_{m+2}), A_{1}(x_{1}),\dots,A_{n}(x_{n}), B(y_{1}),\dots,B(y_{m})) =
\nonumber\\
{1 \over 3!}~g^{(1)}_{bcd}~\sum
T^{\rm conn}(Z_{1}, \xi_{b}(y_{1}), X_{1},Y_{1})~ 
T^{\rm conn}(Z_{2}, \xi_{c}(y_{1}), X_{2},Y_{2})~
T^{\rm conn}(Z_{3}, \xi_{d}(y_{1}), X_{3},Y_{3})
\label{B}
\eea
where the sum 
$
sum
$
runs over the following partitions:
\bea
Z_{1}, Z_{2}, Z_{3} \in Part(\{\xi_{a_{1}},\dots,\xi_{a_{m + 2}}\})
\nonumber\\
X_{1}, X_{2}, X_{3} \in Part(\{A_{1},\dots,A_{n}\})
\nonumber\\
Y_{1}, Y_{2}, Y_{3} \in Part(\{B_{2},\dots,B_{m}\})
\nonumber
\eea
constrained by
\be
{\rm card}(Z_{j}) = {\rm card}(Y_{j}) + 2,~j = 1,2,3. 
\ee
We also have for 
$
n \geq 1
$
\bea
T^{\rm conn}(\xi_{a_{1}}(z_{1}), \dots, \xi_{a_{m+2}}(z_{m+2}), A_{1}(x_{1}),\dots,A_{n}(x_{n}), B(y_{1}),\dots,B(y_{m})) =
\nonumber\\
{1 \over 2!}~g^{(1)}_{ef}~\sum
T^{\rm conn}(Z_{1}, \xi_{e}(x_{1}), X_{1},Y_{1})~ 
T^{\rm conn}(Z_{2}, \xi_{f}(x_{1}), X_{2},Y_{2})
\label{A}
\eea
where the sum 
$
sum
$
runs over the following partitions:
\bea
Z_{1}, Z_{2} \in Part(\{\xi_{a_{1}},\dots,\xi_{a_{m + 2}}\})
\nonumber\\
X_{1}, X_{2} \in Part(\{A_{2},\dots,A_{n}\})
\nonumber\\
Y_{1}, Y_{2} \in Part(\{B_{1},\dots,B_{m}\})
\nonumber
\eea
constrained by
\be
{\rm card}(Z_{j}) = {\rm card}(Y_{j}) + 1,~j = 1,2. 
\ee

We denote formula (\ref{B}) by 
$
(B_{m,n})
$
and formula (\ref{A}) by 
$
(A_{m,n}).
$
\end{thm}
{\bf Proof:} We use a double induction over $m$ and $n$. 

(i) For 
$
m = 0
$
we have only 
$
(A_{0,n})
$
which is Proposition \ref{A(0,n)}. For 
$
m = 1
$
we have 
$
(B_{1,n})
$
which is Proposition \ref{B(1,n)} and 
$
(A_{1,n})
$
which is Proposition \ref{A(1,n)}.

(ii)  First we prove that 
$
(B_{m,n}) \Longrightarrow (A_{m,n})
$
using the same argument as in Proposition \ref{A(1,n)}.

(iii) The induction hypothesis is 
$$
(B_{1,n}),\dots,(B_{m - 1,n}),~(m \geq 2)
$$
and we prove
$
(B_{m,n}).
$
We prove this last formula by induction over $n$.

Then we start the induction proving
$
(B_{m,0}).
$

(iv) Using the equation of motion axiom we have 
\bea
T^{\rm conn}(\xi_{a_{1}}(z_{1}), \dots, \xi_{a_{m+2}}(z_{m+2}), B_{1}(y_{1}),\dots, B_{m}(y_{m})) =
- i~\sum_{j=1}^{m}~D^{F}(\xi_{a_{1}}(z_{1}),\xi_{b}(y_{j}))~
\nonumber\\
T^{\rm conn}(\xi_{a_{2}}(z_{2}), \dots, \xi_{a_{m+2}}(z_{m+2}), B_{1}(y_{1}),\dots,\xi_{b}\cdot B_{j}(y_{j}), \dots, B_{m}(y_{m}))
\eea
and we have two contributions corresponding to
$
j = 1
$
and
$
j > 1.
$
In the first contribution we apply the induction hypothesis 
$
(A_{m - 1,1})
$
and in the second contribution we apply
$
(B_{m-1,1}).
$
After some rearrangements of the terms we get 
$
(B_{m,0}).
$

(v) Now we suppose that 
$$
(B_{m,0}),\dots,(B_{m,n-1})
$$
are valid and we prove
$
(B_{m,n}).
$
As before we have 
\bea
T^{\rm conn}(\xi_{a_{1}}(z_{1}), \dots, \xi_{a_{m+2}}(z_{m+2}), A_{1}(x_{1}, \dots, A_{n}(x_{n}),B_{1}(y_{1}),\dots, B_{m}(y_{m})) =
\nonumber\\
- i~\sum_{j=1}^{m}~D^{F}(\xi_{a_{1}}(z_{1}),\xi_{b}(y_{j}))~
\nonumber\\
T^{\rm conn}(\xi_{a_{2}}(z_{2}), \dots, \xi_{a_{m+2}}(z_{m+2}), 
A_{1}(x_{1}), \dots, A_{n}(x_{n}), B_{1}(y_{1}),\dots,\xi_{b}\cdot B_{j}(y_{j}), \dots, B_{m}(y_{m}))
\nonumber\\
- i~\sum_{j=1}^{n}~D^{F}(\xi_{a_{1}}(z_{1}),\xi_{b}(x_{j}))~
\nonumber\\
T^{\rm conn}(\xi_{a_{2}}(z_{2}), \dots, \xi_{a_{m+2}}(z_{m+2}),
A_{1}(x_{1}), \dots, \xi_{b}\cdot A_{j}(x_{j}), \dots, A_{n}(x_{n}), B_{1}(y_{1}),\dots, B_{m}(y_{m}))
\label{AB}
\eea
The first contribution is analyzed as at (iv): for the term corresponding to
$
j = 1
$
we use 
$
(A_{m-1,n+1})
$
and for the contribution corresponding to
$
j > 1
$
we use 
$
(B_{m-1,n+1}).
$
Finally, for the second contribution from the right hand side of (\ref{AB}) we use the induction hypothesis
$
(B_{m,n-1}).
$
If we cleverly combine the various contributions we obtain 
$
(B_{m,n}).
$
$\qed$

\begin{cor}
(i) Let 
$
A_{1},\dots,A_{n}
$
trilinear monomials. Then we have for the connected tree contributions
$
T^{\rm conn}_{(0)}
$
the following formula for
$
n \geq 4
$
\be
T^{\rm conn}_{(0)}(A_{1}(x_{1}),\dots,A_{n}(x_{n}))
= {1 \over 3!}~g^{(1)}_{bcd}~\sum : T^{\rm conn}_{(0)}(I_{1},\xi_{b}(x_{1}))~T^{\rm conn}_{(0)}(I_{2},\xi_{c}(x_{1}))~
T^{\rm conn}_{(0)}(I_{3},\xi_{d}(x_{1})):
\ee
where the sum runs over the partitions of the set 
$
A_{2},\dots,A_{n}.
$

(ii) Let 
$
A_{1}
$
be a bilinear Wick monomial and 
$
A_{2},\dots,A_{n}
$
trilinear monomials. Then we have  for 
$
n \geq 3
$
\be
T^{\rm conn}_{(0)}(A_{1}(x_{1}),\dots,A_{n}(x_{n}))
= {1 \over 2!}~g^{(1)}_{ef}~\sum : T^{\rm conn}_{(0)}(I_{1},\xi_{e}(x_{1}))~T^{\rm conn}_{(0)}(I_{2},\xi_{f}(x_{1})):
\ee
where the sum runs over the partitions of the set 
$
A_{2},\dots,A_{n}.
$
\label{amplitude}
\end{cor}
{\bf Proof:} We combine Wick theorem \ref{wick-thm}  and the preceding theorem.
$\qed$

Now we are ready to study the so-called Berends-Giele currents \cite{BG2}. We observe from (i) of the preceding corollary that the 
connected part of the tree contributions to the chronological products 
$
T^{\rm conn}_{(0)}(B_{1}(y_{1}),\dots,B_{n}(y_{n}))
$
can be expressed in terms of chronological products of the type
$
T^{\rm conn}_{(0)}(\xi_{a}(z),B_{1}(y_{1}),\dots,B_{m}(y_{m})),~m < n.
$
These are the Berends-Giele currents in coordinate space. There is a special entry linear in the fields which corresponds to the 
special line of \cite{BG2}; The field
$
\xi_{a}(z)
$
is taken to be off-shell in this reference.
First we give the recursion formula for such currents which is a coordinates version of the recursion formula of \cite{BG2}.
\begin{prop}
Suppose that 
\be
B_{j} = {1 \over 3}~ g^{(j)}_{bcd}~\xi_{b}\xi_{c}\xi_{d},
\ee
with 
$
g^{(j)}_{bcd}
$
completely symmetric. Then the following recursion formulas are valid for
$
n \geq 3
$
\bea
T^{\rm conn}_{(0)}(\xi_{a_{0}}(z_{0}),B_{1}(y_{1}),\dots,B_{n}(y_{n})) = 
\nonumber\\
- {i \over 2}~\sum_{j=1}^{m}~g^{(j)}_{bcd}~
D^{F}(\xi_{a_{0}}(z_{0}),\xi_{b}(y_{j}))~
\sum_{I_{1},I_{2}}~:T^{\rm conn}_{(0)}(\xi_{c}(y_{j}),I_{1})~T^{\rm conn}_{(0)}(\xi_{d}(y_{j}),I_{2}):
\eea
where the sum 
$
\sum_{I_{1},I_{2}}
$
goes over the partitions of
$
\{ B_{1}(y_{1}),\dots,\hat{j},\dots,B_{n}(y_{n})\}.
$
\label{BG-coordinate}
\end{prop}
{\bf Proof:} First we use the equation of motion formula:
\bea
T^{\rm conn}_{(0)}(\xi_{a_{0}}(z_{0}),B_{1}(y_{1}),\dots,B_{n}(y_{n})) = 
\nonumber\\
- i~\sum_{j=1}^{m}~D^{F}(\xi_{a_{0}}(z_{0}),\xi_{b}(y_{j}))
T^{\rm conn}_{(0)}(B_{1}(y_{1}),\dots,\xi_{b}\cdot B_{j}(y_{j}),\dots,B_{n}(y_{n}))
\eea
and then we use for the chronological product from the right hand side part (ii) of the preceding Corollary with 
$
A_{1} \rightarrow \xi_{b}\cdot B_{j}
$
and
$
A_{2},\dots,A_{n} \rightarrow B_{1},\dots,\hat{j},\dots,B_{n}.
$
$\qed$

We should compare the preceding formula with formula (52) of \cite{Di}, more precisely the first line and notice a similar structure.

We now go to momentum space. A typical experiment is described by the matrix element 
\be
< \tilde{\xi}^{(+)}_{a_{1}}(p_{1}) \cdots \tilde{\xi}^{(+)}_{a_{n}}(p_{n})\Omega, 
T^{\rm conn}_{(0)}(\xi_{a}(z),B_{1}(y_{1}),\dots,B_{m}(y_{m}))\Omega>
\label{typical}
\ee
which creates $n$ particles from the vacuum. Here
$
\tilde{\xi}_{a}(p)
$
are the Fourier transforms of the fields and 
$
\tilde{\xi}^{(\pm)}_{a}(p)
$
are the positive and negative parts.

We have
\begin{prop}
The preceding expression can be non-zero only for 
$
n = m + 1.
$
\end{prop}
{\bf Proof:} From Wick theorem and the first Proposition \ref{t-conn}. Indeed, from Wick theorem we have 
\bea
T^{\rm conn}_{(0)}(\xi_{a}(z),B_{1}(y_{1}),\dots,B_{m}(y_{m})) \sim
\nonumber\\
\sum < \Omega, T^{\rm conn}_{(0)}(\xi_{a}(z),B^{\prime}_{1}(y_{1}),\dots,B^{\prime}_{m}(y_{m}))\Omega>
:B^{\prime\prime}_{1}(y_{1})\dots B^{\prime\prime}_{m}(y_{m}):
\nonumber
\eea
where 
$
B = B^{\prime}B^{\prime\prime}
$
is a factorization of $B$ in Wick submonomials. Suppose that in the set of Wick monomials 
$
B^{\prime}_{1}(y_{1}),\dots,B^{\prime}_{m}(y_{m})
$
we have 
$
m_{j} 
$
elements with 
$
\rho = j,~j =1,2,3
$
i.e. with 
$
1, 2, 3
$
factor fields. Then in 
$
:B^{\prime\prime}_{1}(y_{1}),\dots,B^{\prime\prime}_{m}(y_{m}):
$
we have
$
2 m_{1} + m_{2}
$
factors so the matrix element is non-zero only {\it iff}
$
n = 2 m_{1} + m_{2}.
$
Using Proposition \ref{t-conn} we get the equality from the statement.
$\qed$

In momentum space we want to obtain recursion formulas for the expressions
\bea
T^{B_{1},\dots,B_{m}}_{a_{0};a_{1},\dots,a_{n}}(p_{0};p_{1},\dots,p_{n})
= {1 \over (2\pi)^{2}}~\int dy_{1}\cdots dy_{m} dz_{0}~e^{- i p_{0}\cdot z_{0}}~
\nonumber\\
< \tilde{\xi}^{(+)}_{a_{1}}(p_{1}) \cdots \tilde{\xi}^{(+)}_{a_{n}}(p_{n})\Omega, 
T^{\rm conn}_{(0)}(\xi_{a}(z),B_{1}(y_{1}),\dots,B_{m}(y_{m}))\Omega>
\label{Tn}
\eea
and we need to consider only the case
$
n = m + 1
$
as we have proved above. 

We will give a recursion formula for these objects. First we need a technical lemma.
\begin{lemma}
Let 
\be
A = \prod_{j=1}^{n} \xi^{(-)}_{a_{j}}(z_{j}),\quad 
B_{k} = \prod_{j \in P_{k}} \xi^{(+)}_{b_{j}}(y_{j}),~(k = 1,2)
\ee
with 
$
P_{1} \cup P_{2} = \{1,\dots,n\},~ P_{1} \cap P_{2} = \emptyset,~ P_{1} \not= \emptyset,~P_{2} \not= \emptyset.
$
Then we have
\be
<\Omega, A B_{1} B_{2}\Omega> = \sum <\Omega, A_{1} B_{1}\Omega>~<\Omega, A_{2} B_{2}\Omega>
\ee
where the sum goes over all factorizations 
$
A = A_{1}A_{2}
$
into submonomials such that
$
\rho(A_{j}) = \rho(B_{j}),~j = 1,2.
$
\end{lemma}
{\bf Proof:} We use induction over $n$. For 
$
n = 2
$
the result is easy to obtain. We suppose that we have proved the result for 
$
2,\dots, n
$
and consider the case 
$
n + 1
$.
We take
$
A =  A_{0} \xi^{(-)}_{a_{n+1}}(z_{n+1})
$
where
$
A_{0} = \prod_{j=1}^{n} \xi^{(-)}_{a_{j}}(z_{j})~
$
and compute
\bea
<\Omega, A_{0}~\xi^{(-)}_{a_{n+1}}(z_{n+1}) B_{1}B_{2}\Omega> =
\nonumber\\
<\Omega, A_{0}~[\xi^{(-)}_{a_{n+1}}(z_{n+1}), B_{1}]B_{2}\Omega>~
+ <\Omega, A_{0}~B_{1}~[\xi^{(-)}_{a_{n+1}}(z_{n+1}), B_{2}]\Omega>
\nonumber
\eea
The commutators are sums of expressions of the type
$
\prod_{j} \xi^{(+)}_{b_{j}}(y_{j})
$
with one less factor $\xi$ and some numerical coefficients. So, we can apply the induction hypothesis and obtain
\bea
<\Omega, A B_{1} B_{2}\Omega> 
\nonumber\\
= \sum <\Omega, A_{1}~[\xi^{(-)}_{a_{n+1}}(z_{n+1}), B_{1}]\Omega>~<\Omega, A_{2} B_{2}\Omega>
\nonumber\\
+ \sum <\Omega, A_{1} B_{1}\Omega>~<\Omega, A_{2}~[\xi^{(-)}_{a_{n+1}}(z_{n+1}), B_{2}]\Omega>
\nonumber
\eea
where the first sum is restricted to
$
\rho(A_{1}) = \rho(B_{1}) - 1,~\rho(A_{2}) = \rho(B_{2})
$
and the second sum to
$
\rho(A_{1}) = \rho(B_{1}),~\rho(A_{2}) = \rho(B_{2}) - 1.
$
We can rewrite the preceding formula as
\bea
<\Omega, A B_{1} B_{2}\Omega> 
\nonumber\\
= \sum <\Omega, A_{1}~\xi^{(-)}_{a_{n+1}}(z_{n+1}), B_{1}\Omega>~<\Omega, A_{2} B_{2}\Omega>
\nonumber\\
+ \sum <\Omega, A_{1} B_{1}\Omega>~<\Omega, A_{2}~\xi^{(-)}_{a_{n+1}}(z_{n+1}) B_{2}\Omega>
\nonumber
\eea
and observe that the factorizations 
$
A = A_{1}A_{2}
$
into submonomials such that
$
\rho(A_{j}) = \rho(B_{j}),~j = 1,2
$
are of two types
$
A = (A_{1}\xi^{(-)}_{a_{n+1}}(z_{n+1})) A_{2}
$
such that
$
\rho(A_{1}) = \rho(B_{1}) - 1,~\rho(A_{2}) = \rho(B_{2})
$
and
$
A = A_{1} (A_{2}\xi^{(-)}_{a_{n+1}}(z_{n+1}))
$
with
$
\rho(A_{1}) = \rho(B_{1}),~\rho(A_{2}) = \rho(B_{2}) - 1.
$
They correspond to the two sums above.
$\qed$

Using this lemma we obtain from Proposition \ref{BG-coordinate}:
\begin{prop}
Suppose that 
\be
B_{j} = {1 \over 3}~ g^{(j)}_{bcd}~\xi_{b}\xi_{c}\xi_{d},
\ee
with 
$
g^{(j)}_{bcd}
$
completely symmetric. Then the following recursion formulas are valid for 
$
n \geq 4
$
\bea
< \tilde{\xi}^{(+)}_{a_{1}}(p_{1}) \cdots \tilde{\xi}^{(+)}_{a_{n}}(p_{n})\Omega,
T^{\rm conn}_{(0)}(\xi_{a_{0}}(z_{0}),B_{1}(y_{1}),\dots,B_{m}(y_{m})) \Omega> = 
\nonumber\\
- {i \over 2}~\sum_{j=1}^{m}~g^{(j)}_{bcd}~
D^{F}(\xi_{a_{0}}(z_{0}),\xi_{b}(y_{j}))~\sum_{I_{1},I_{2}}~\sum_{P_{1},P_{2}}
< \Omega, \prod_{j \in P_{1}} \tilde{\xi}^{(-)}_{a_{j}}(p_{j})~T^{\rm conn}_{(0)}(\xi_{c}(y_{j}),I_{1})\Omega>
\nonumber\\
< \Omega, \prod_{j \in P_{2}} \tilde{\xi}^{(-)}_{a_{j}}(p_{j})~~T^{\rm conn}_{(0)}(\xi_{d}(y_{j}),I_{2})\Omega>
\eea
where the sum 
$
\sum_{I_{1},I_{2}}
$
goes over the partitions of
$
B_{1}(y_{1}),\dots,\hat{j},\dots,B_{m}(y_{m})
$
and the sum
$
\sum_{P_{1},P_{2}}
$
over partitions of
$
\{1,\dots,n\}
$
such that
$
card(P_{j}) = card(I_{j}) + 1,\quad j = 1,2.
$
\end{prop}
Finally we have
\begin{thm}
The expressions (\ref{Tn}) are of the form 
\be
T^{B_{1},\dots,B_{m}}_{a_{0};a_{1},\dots,a_{n}}(p_{0};p_{1},\dots,p_{n}) = \delta(\sum_{j = 0}^{n} p_{j})
A^{B_{1},\dots,B_{m}}_{a_{0};a_{1},\dots,a_{n}}(p_{1},\dots,p_{n})
\ee
and the following recursion relations are true
\bea
A^{B_{1},\dots,B_{m}}_{a_{0};a_{1},\dots,a_{n}}(p_{0};p_{1},\dots,p_{n}) =
- {i \over 2}~(2\pi)^{6}~\sum_{j=1}^{m}~g^{(j)}_{bcd}~\tilde{D}^{F}_{a_{0}b}(p_{0})~
\sum_{I_{1},I_{2}}~\sum_{P_{1},P_{2}}~A^{I_{1}}_{c,{\cal A}_{1}}({\cal P}_{1})~A^{I_{2}}_{d,{\cal A}_{2}}({\cal P}_{2})
\label{An}
\eea
where the sums are as above and we have denoted for simplicity
$
p_{0} \equiv - \sum_{k=1}^{n}p_{k}
$
and
\bea
{\cal A}_{1} = \{ a_{j} \}_{j \in P_{1}},~{\cal P}_{1} = \{ p_{j} \}_{j \in P_{1}},\quad
{\cal A}_{2} = \{ a_{j} \}_{j \in P_{2}},~{\cal P}_{2} = \{ p_{j} \}_{j \in P_{2}}.
\nonumber
\eea
\label{bg-recursion}
\end{thm}
{\bf Proof:} We integrate over 
$
y_{1},\dots,y_{n}
$
the formula from the preceding Proposition. The integral over
$
y_{k} \not= y_{j}
$
corresponding to
$
B_{k} \in I_{1}
$
and
$
B_{k} \in I_{2}
$
respectively, are producing, essentially, the two factors $A$ from the right hand side. The integral over 
$
y_{j}
$
and
$
z_{0}
$
are producing the overall $\delta$ factor multiplied by the propagator
$
\tilde{D}^{F}_{a_{0}b}(p_{0}).
$
$\qed$

\newpage
\section{Yang-Mills Fields\label{ym}}

First, we can generalize the preceding formalism to the case when some of the scalar fields
are odd Grassmann variables. One simply insert everywhere the Koszul sign. The next generalization is to arbitrary vector and spinorial
fields. If we consider for instance the Yang-Mills interaction Lagrangian corresponding to pure QCD \cite{algebra} then the jet variables 
$
\xi_{a}, a \in \Xi
$
are
$
(v^{\mu}_{A}, u_{A}, \tilde{u}_{A}),~A = 1,\dots,r
$
where 
$
v^{\mu}_{A}
$
are Grassmann even and 
$
u_{A}, \tilde{u}_{A}
$
are Grassmann odd variables. 

The interaction Lagrangian is determined by gauge invariance. Namely we define the {\it gauge charge} operator by
\be
d_{Q} v^{\mu}_{A} = i~d^{\mu}u_{A},\qquad
d_{Q} u_{A} = 0,\qquad
d_{Q} \tilde{u}_{A} = - i~d_{\mu}v^{\mu}_{A},~A = 1,\dots,r
\ee
where 
$
d^{\mu}
$
is the formal derivative. The gauge charge operator squares to zero:
\be
d_{Q}^{2} \simeq  0
\ee
where by
$
\simeq
$
we mean, modulo the equation of motion. Now we can define the interaction Lagrangian by the relative cohomology relation:
\be
d_{Q}T(x) \simeq {\rm total~divergence}.
\ee
If we eliminate the corresponding coboundaries, then a tri-linear Lorentz covariant 
expression is uniquely given by
\bea
T = f_{ABC} \left( {1\over 2}~v_{A\mu}~v_{B\nu}~F_{C}^{\nu\mu}
+ u_{A}~v_{B}^{\mu}~d_{\mu}\tilde{u}_{C}\right)
\label{Tint}
\eea
where
\be
F^{\mu\nu}_{A} \equiv d^{\mu}v^{\nu}_{A} - d^{\nu}v^{\mu}_{A}, 
\quad \forall a = 1,\dots,r
\ee 
and 
$
f_{ABC}
$
are real and completely anti-symmetric. (This is the tri-linear part of the usual QCD interaction Lagrangian from classical field theory.)

Then we define the associated Fock space by the non-zero $2$-point distributions are
\bea
<\Omega, v^{\mu}_{A}(x_{1}) v^{\nu}_{B}(x_{2})\Omega> = 
i~\eta^{\mu\nu}~\delta_{AB}~D_{0}^{(+)}(x_{1} - x_{2}),
\nonumber \\
<\Omega, u_{A}(x_{1}) \tilde{u}_{B}(x_{2})\Omega> = - i~\delta_{AB}~D_{0}^{(+)}(x_{1} - x_{2}),
\nonumber\\
<\Omega, \tilde{u}_{A}(x_{1}) u_{B}(x_{2})\Omega> = i~\delta_{AB}~D_{0}^{(+)}(x_{1} - x_{2}).
\label{2-massless-vector}
\eea
and construct the associated Wick monomials. Then the expression (\ref{Tint}) gives a Wick polynomial 
$
T^{\rm quant}
$
formally the same, but: 
(a) the jet variables must be replaced by the associated quantum fields; (b) the formal derivative 
$
d^{\mu}
$
goes in the true derivative in the coordinate space; (c) Wick ordering should be done to obtain well-defined operators. We also 
have an associated {\it gauge charge} operator in the Fock space given by
\bea
~[Q, v^{\mu}_{A}] = i~\partial^{\mu}u_{A},\qquad
\{ Q, u_{A} \} = 0,\qquad
\{Q, \tilde{u}_{A}\} = - i~\partial_{\mu}v^{\mu}_{A}
\nonumber \\
Q \Omega = 0.
\label{Q-vector-null}
\eea

Then it can be proved that
$
Q^{2} = 0
$
and
\be
~[Q, T^{\rm quant}(x) ] = {\rm total~divergence}
\ee
where the equations of motion are automatically used because the quantum fields are on-shell.
From now on we abandon the super-script {\it quant} because it will be obvious from the context if we refer 
to the classical expression (\ref{Tint}) or to its quantum counterpart.

Next, we notice that we can write (\ref{Tint})  in the form (\ref{tri}):
\be
T = {1\over 3!}~g_{pqr}~\xi_{p}~\xi_{q}~\xi_{r}
\label{T-gen}
\ee
with
$
\xi_{p}
$
jet variables and
\be
g_{pqr} \equiv g(\xi_{p},\xi_{q},\xi_{r})
\ee
having Grassmann permutation symmetries. For the QCD Lagrangian from above we have the non-zero entries:
\bea
g(v_{A}^{\mu},v_{B}^{\nu},F_{C}^{\rho\sigma}) = {1\over 2}~(\eta^{\mu\sigma}~\eta^{\nu\rho} - \eta^{\nu\sigma}~\eta^{\mu\rho})~f_{ABC}
\nonumber\\
g(u_{A},v_{B}^{\mu},\tilde{u}_{C;\nu}) = \delta^{\mu}_{\nu}~f_{ABC}
\eea
and the rest of the non-zero expressions
$
g_{pqr}
$
obtained by permutations, taking into account the Grassmann parity sign.

It is known that if we construct the chronological products in the second order, then gauge invariance can be saved if we impose that 
the constants
$
f_{ABC}
$
are verifying Jacobi identity and add a finite renormalization of the type  \cite{Sc2}
\be
N(T(x),T(y)) = \delta(x - y)~N(x), \quad N \equiv {i \over 2}~f_{ABE}~f_{CDE}~v_{A}^{\mu}v_{B}^{\nu}v_{C\mu}v_{D\nu}.
\ee
The Wick polynomial $N$ is the quadri-linear part of the usual QCD interaction Lagrangian from classical field theory and it is the
justification for the quadri-linear vertexes from \cite{BG2} (see more precisely \cite{Di}, the second line of formula (52)).

We notice that we can get rid of the finite renormalization if one redefines:
\bea
T( v_{A\mu;\nu}(x), v_{B\rho;\sigma}(y) ) = D^{F}( v_{A\mu;\nu}(x), v_{B\rho;\sigma}(y)  ) =
\nonumber\\
- i~\delta_{AB}~\eta_{\mu\rho}~\partial_{\nu}\partial_{\sigma}D^{F}(x - y) 
+ c~\eta_{\mu\rho}~\eta_{\nu\sigma}~\delta_{AB}~\delta(x - y)
\label{FF}
\eea
with a clever choice
$
c = {i\over 2}.
$
The first term in the right hand side follows from the causal splitting of formula (\ref{2-point-der}). The second term is possible because it does not change 
the order of singularity and the covariance properties of the chronological product.
So, we this clever trick we do not need the quadri-linear vertexes of \cite{BG2}.

Now we try to make the connection with the momentum space formulas i.e we consider expressions of the type (\ref{Tn}). In 
the case of gauge models we should pay attention to the choice of the physical external states of the type
$
\tilde{\xi}^{(+)}_{a_{1}}(p_{1}) \cdots \tilde{\xi}^{(+)}_{a_{n}}(p_{n})\Omega.
$
It can be proved \cite{cohomology} that the physical states of the Yang-Mills setting are of the form 
$
\epsilon(p_{1}, \pm)\cdot\tilde{v}_{A_{1}}(p_{1}) \cdots \epsilon(p_{n}, \pm)\cdot\tilde{v}_{A_{n}}(p_{n})\Omega
$
where the expressions
$
\epsilon(p_{j}, \pm)
$
are selecting the two polarizations of a gluon state. (It means that we can ignore all tree graphs with ghost lines).
It follows that the basic Feynman amplitudes are of the form
\be
A(A_{0},\epsilon_{0},p_{0};A_{1},\epsilon_{1},p_{1};\dots;A_{n},\epsilon_{n},p_{n})
\label{amplitude-ym}
\ee
where
$
\epsilon_{j} = \pm.
$
We note that we have taken in (\ref{Tn})
$
B_{1} = \dots = B_{m} = T 
$
where $T$ is the interaction Lagrangian given by (\ref{Tint}).

The idea from \cite{BG2} is to separate the color dependence; mathematically it means to find an orthogonal basis in the space 
of tensors
$
t^{(K)}_{A_{0}\dots A_{n}}
$
and write
\be
A(A_{0},\epsilon_{0},p_{0};A_{1},\epsilon_{1},p_{1};\dots;A_{n},\epsilon_{n},p_{n}) = \sum 
t^{(K)}_{A_{0}\dots A_{n}}~A_{K}(\epsilon_{0},p_{0};\epsilon_{1},p_{1};\dots;\epsilon_{n},p_{n}).
\ee
If such a writing would be possible we would used it in theorem \ref{bg-recursion} and determine a recursion relation for the reduced 
amplitude of the type
$
A_{K}(\epsilon_{0},p_{0};\epsilon_{1},p_{1};\dots;\epsilon_{n},p_{n}).
$
However to find such a basis is not elementary. We give some details. By
$
A \rightarrow F_{A}
$ 
we mean the adjoint representations of the Lie algebra associated to 
$
f_{ABC}
$
\be
(F_{A})_{BC} = - f_{ABC}.
\ee

We will assume that the Lie algebra associated to the structure constants
$
f_{ABC}
$
is semi-simple so the Killing-Cartan form
$
g_{AB} = f_{ACD}~f_{BDC}
$
can be chosen 
$
g_{AB} = \delta_{AB}.
$
It follows that the 
$
r \times r
$
matrices
$
F_{A},~A = 1,\dots,r 
$
are linear independent. In this case we can prove the following ``completness" relation:
\begin{lemma}
Suppose that
$
X, Y
$
are in the linear span of
$
F_{A},~A = 1,\dots,r. 
$
Then the following formula is true 
\be
f_{ABC}~Tr(F_{B}X)~Tr(F_{C}Y) = Tr(F_{A}[X,Y])
\ee
where we sum over the dummy indices.
\end{lemma}
{\bf Proof:}
We introduce in 
$
{\cal V} \equiv {\rm Span}(F_{A})_{A= 1,\dots,r}
$
the scalar product
\be
(X,Y) = Tr(XY).
\ee
In particular
\be
(F_{A},F_{B}) = \delta_{AB}.
\ee
Because 
$
X, Y \in {\cal V}
$
by hypothesis, we have the writings
\be
X = x_{B}~F_{B},\quad Y = y_{B}~F_{B}
\ee
from where
\be
[ X,Y] = f_{ABC}~x_{B}~y_{C}~F_{A}.
\ee
If we substitute in the relation from the statement we obtain an identity.
$
\qed
$

Next we define the iterated commutators \cite{EKMZ}
$
C(A_{1},\dots,A_{n})
$
through the recursion relations:
\bea
C(\emptyset) = 1,~C(A_{1}) = F_{A_{1}}
\nonumber\\
C(A_{1},\dots,A_{n-1}) = [ C(A_{1},\dots,A_{n-2}), F_{A_{n-1}}].
\label{iterated}
\eea
and note that they belong to 
$
{\cal V}.
$

We now prove the following
\begin{prop}
The color factors the amplitudes (\ref{amplitude-ym}) are of the form
$$
Tr(C(A_{1},\dots,A_{n})F_{A_{0}}).
$$
\end{prop}
{\bf Proof:} The first non-trivial case is
$
n = 2
$
(and
$
m = 1
$)
the expression (\ref{typical}) is proportional to 
$
g_{a_{0}a_{1}a_{2}}
$
so for the particular case of QCD we get the factor
$
f_{A_{0}A_{1}A_{2}} \sim Tr((C(A_{1},A_{2})F_{A_{0}})
$.

If we suppose that the assertion from the statement is true for 
$
2,\dots,n - 1
$
we have for $n$ from (\ref{An}) that the color factor is of the form 
$$
f_{BCD}~Tr(C(A_{1},\dots,A_{k})F_{B})~Tr(C(A_{k+1},\dots,A_{n})F_{C}).
$$
If we apply the preceding lemma we obtain the color factor
$$
Tr([C(A_{1},\dots,A_{k}), C(A_{k+1},\dots,A_{n})]F_{A}).
$$
But the comutator of two iterated commutator is a sum of iterated commutators, so we obtain the result from the statement.
$
\qed
$

So it follows that the Feynman amplitudes are sums of the type
\be
A = \sum Tr(C(A_{\sigma(1)},\dots,A_{\sigma(n)})F_{A_{0}})~A_{\sigma}.
\ee
If we explicitate the commutators we arrive at the fromula from \cite{BG1}:
\be
A_{A_{0};A_{1}\dots A_{n}} = \sum Tr(F_{A_{\sigma(0)}} \dots F_{A_{\sigma(n)}})~A_{\sigma}
\label{non-unique}
\ee

However the expressions 
$
Tr(F_{A_{0}} \dots F_{A_{n}})
$
above are not linear independent. They verify many identities presented in \cite{BG1} and  \cite{KK}.
If one takes into account all these identities one can find, at least for the Lie algebra
$
su(N),
$
an orthogonal basis, up to terms of order
$
1/ N^{2}
$
- see \cite{MP} formula (3.3).

Next we have 
\begin{prop}
The following formula is valid:
\be
(F_{A_{2}}\dots F_{A_{n-1}})_{A_{1}A_{n}} = Tr( C(A_{1},\dots,A_{n-2})~[ F_{A_{n-1}}, F_{A_{n}}]) =
Tr(C(A_{1},\dots,A_{n-2}, A_{n-1}) F_{A_{n}})
\ee
\end{prop}
{\bf Proof:} By induction. For 
$
n = 3
$
the assertion is immediate. If the formula from the statement is true then we have
\bea
(F_{A_{2}}\dots F_{A_{n}})_{A_{1}A_{n+1}} = (F_{A_{2}}\dots F_{A_{n-1}})_{A_{1}B}~(F_{A_{n}})_{BA_{n+1}}
\nonumber\\
= Tr( C(A_{1},\dots,A_{n-2}) [ F_{A_{n-1}}, F_{B}])~f_{BA_{n}A_{n+1}}
\nonumber\\
= Tr( C(A_{1},\dots,A_{n-2}) [ F_{A_{n-1}}, [F_{A_{n}}, F_{A_{n+1}}]])
\nonumber\\
= Tr( C(A_{1},\dots,A_{n-2}) F_{A_{n-1}}[F_{A_{n}}, F_{A_{n+1}}])
- Tr( C(A_{1},\dots,A_{n-2}) [F_{A_{n}}, F_{A_{n+1}}] F_{A_{n-1}}).
\nonumber
\eea
We have used the induction hypothesis in the second equality. Now, in the very last term from above 
we use the cyclic property of the trace and immediately obtain the formula from the statement for 
$
n \rightarrow n + 1
$
and this finishes the induction.
$\qed$

If we use this formula we get color factors of the form
$
(F_{A_{2}}\dots F_{A_{n-1}})_{A_{1}A_{n}}
$;
in \cite{DDM} (see also \cite{EKMZ}) the linear independence for this basis was proved for 
$
A_{1}, A_{n}
$
fixed; however orthogonality is lacking. From the result of \cite{DDM} we can obtain the result of \cite{BG1}. 
For an orthogonal basis see \cite{Z}. 

\section{Conclusions}
We have proved that the factorization formula is very simple in the coordinate space; it is based on Proposition \ref{BG-coordinate} and 
Corollary \ref{amplitude}. To go in the momentum space requires some computations and the result is given in Theorem \ref{bg-recursion}.

One can easily particularize the formulas for the Yang-Mills case. However, to strip the amplitudes of the color factors one needs 
an linear independent and orthogonal basis in the space of tensors associated to the Lie algebra with some special
tensor properties and this is not an elementary problem.

We stress in the end the fact that our factorization formula is similar, but not identical to the factorization formula of \cite{BG2}.
In our opinion, it remains an interesting problem to derive the factorization formula of \cite{BG2} using only well defined 
mathematical objects, namely the chronological products.

\end{document}